\documentclass[fleqn,twoside]{article}
\usepackage{espcrc2}

\usepackage{graphicx}
\usepackage[figuresright]{rotating}

\newcommand{\AmS}{{\protect\the\textfont2
A\kern-.1667em\lower.5ex\hbox{M}\kern-.125emS}}
\newcommand\cpc[3]{{\it Comput. Phys. Commun. }{\bf #1} (#2) #3}
\newcommand\npb[3]{{\it Nucl. Phys. }{\bf B #1} (#2) #3}

\newcommand\plb[3]{{\it Phys. Lett. }{\bf B #1} (#2) #3}

\newcommand\sjnp[3]{{\it Sov. J. Nucl. Phys. }{\bf #1} (#2) #3}
\newcommand\jetp[3]{{\it Sov. Phys. JETP }{\bf #1} (#2) #3}
\newcommand\zpc[3]{{\it Z. Physik }{\bf C #1} (#2) #3}

\title{The $\tt BABAYAGA$ event generator}

\author{C.M. Carloni Calame
\address[PVINFN]{Istituto Nazionale di Fisica Nucleare, Sezione di Pavia, 
Via A. Bassi, 6 - 27100 Pavia, Italy}
\address[DFNT]{Dipartimento di Fisica Nucleare e Teorica,
Via A. Bassi, 6 - 27100 Pavia, Italy},
G. Montagna \addressmark[DFNT] \addressmark[PVINFN],
O. Nicrosini \addressmark[PVINFN] \addressmark[DFNT] and 
F. Piccinini \addressmark[PVINFN] \addressmark[DFNT].}
\begin{document}
\begin{abstract}
\noindent 
The program $\tt BABAYAGA$ is an event generator for QED processes at flavour
factories, mainly intended for luminosity measurement of $e^+e^-$ colliders
in the center of mass range 1-10 GeV.
Recently, the $\pi^+\pi^-$ channel has been added as well. 
The relevant (photonic) radiative corrections are simulated by means of a
Parton Shower in QED. The theoretical precision of the approach is estimated
and some phenomenological results are discussed.
\vspace{1pc}
\end{abstract}
\maketitle

\section{Introduction}
The precise determination of the machine luminosity is an important ingredient
for the successful achievement of the physics programme at the $e^+e^-$
colliders running with center of mass energy in the range of the low lying
hadronic resonances (1 - 10 GeV).

One of the most important challenges is the precise measure of
the $R$ ratio, by means of the energy scan or the radiative return method.
The aim is to reduce the theoretical error on the hadronic contribution to the
vacuum polarization, which will reflect on the error of the anomalous magnetic
moment of the muon $a_\mu$ and the QED coupling constant at the $Z$ peak
$\alpha_{QED}(M_Z^2)$ \cite{fj}.
Actually, the $R$ measurement is going to become a precision measurement 
\cite{nsk,incagli} and
it will give a stringent test of the Standard Model predictions.

In this perspective, a precise knowledge of the cross section for QED
processes for luminometry, expecially for Bhabha scattering, and their Monte
Carlo (MC) simulation are crucial. The theoretical accuracy should be better
than $0.5\%$.

\section{The theoretical approach in $\tt BABAYAGA$}
$\tt BABAYAGA$ \cite{ournpb,myplb} is a Monte Carlo event generator for the
simulation of QED and $\pi^+\pi^-$ processes at flavour factories.

In order to achieve an accurate prediction of the cross sections of the QED 
processes, the relevant radiative correction (RC) have to be accounted for,
and, in particular, photon radiation effects must be included in the
calculation. 

\subsection{The QED Structure Functions}
The corrected cross section and the event generation can be obtained
according to the following master formula (see \cite{ournpb} and references
therein):
\begin{eqnarray}
&& \sigma(s)=\int d x_1 d x_2 d y_1 d y_2\int d \Omega \times \nonumber\\
&& D(x_1,Q^2)D(x_2,Q^2)D(y_1,Q^2)D(y_2,Q^2)\times \nonumber\\
&& \frac{ d \sigma_0(x_1x_2s)}{d \Omega}\Theta(cuts)
\label{sezfs}
\end{eqnarray}
The previous equation is used in $\tt BABAYAGA$ for the generation of the
processes $e^+e^-\to e^+e^-(n\gamma)$, $\to \mu^+\mu^-(n\gamma)$, 
$\to\gamma\gamma(n\gamma)$ and $\to \pi^+\pi^-(n\gamma)$. The last channel has
been recently implemented and, in this case, only photons coming from initial
state are accounted for.
 
In eq. \ref{sezfs}, the Born-like cross section for the process
$d\sigma_0/d\Omega$ is convoluted with the QED non-singlet
Structure Functions (SF's) $D(x,Q^2)$, which are the solution of the 
Dokshitzer-Gribov-Lipatov-Altarelli-Parisi (DGLAP)
\cite{dglapref} equation in QED
\begin{eqnarray}
&& Q^2\frac{\partial}{\partial Q^2}D(x,Q^2)=\frac{\alpha}{2\pi}
\int_x^{1}\frac{dy}{y} P_+(y) D(\frac{x}{y},Q^2)\nonumber \\
&& P_+(x)=\frac{1+x^2}{1-x}-\delta(1-x)\int_0^1 dt P(t)
\label{qeddglap}
\end{eqnarray}
where $P_+(x)$ is the regularized Altarelli-Parisi vertex.
The SF's in eq. \ref{sezfs} account for photon radiation emitted
by both initial-state (ISR) and final-state (FSR) fermions.
The SF's allow to include the universal virtual
and real photonic corrections, up to all orders of perturbation theory.
Strictly speaking, radiation in eq. \ref{sezfs} is collinear to the emitting
fermions.

It is possible to go beyond the collinear approximation by solving the DGLAP
equation by means of a MC algorithm, the so-called Parton Shower algorithm
(PS), which is discussed in the next section.

\subsection{The Parton Shower algorithm}
\label{PSalgo}
The Parton Shower is a MC algorithm which solves exactly the DGLAP equation
(see references in \cite{ournpb}) and it originates from an iterative solution
of eq. \ref{qeddglap}
\begin{eqnarray}
&&D(x,Q^2)=\Pi(Q^2,m^2)\delta(1-x) \nonumber\\
&&+\frac{\alpha}{2\pi}\int_{m^2}^s\Pi (Q^2,s')\frac{d s'}{s'}\Pi (s',m^2)\times
\nonumber\\
&&\times\int_0^{x_+}dyP(y)\delta (x-y) + \nonumber \\
&& + \ 2 \ photons +\cdots
\label{APitera}
\end{eqnarray}
In eq. \ref{APitera}, 
$\Pi(Q^2,m^2)=e^{-\frac{\alpha}{2\pi}\log(\frac{Q^2}{m^2})
\int_\epsilon^1 P(x)}$ is the Sudakov Form Factor, where the
soft and virtual RC are exponentiated. The $Q^2$ scale is not {\em a priori} 
fixed, but it has to be chosen in order to reproduce as close as possible the
leading
RC contributions coming from the exact order $\alpha$ calculation. For example,
for the Bhabha scattering, a reasonable choice is $Q^2=st/ue$, where $s$,$t$
and $u$ are the usual Mandelstam variables and $e$ is the Euler's number.

The main advantage of this solution is that it keeps track of the number of
the emitted photons and that their kinematics can be approximately recovered.
Therefore, the whole branching event can be generated and the multi-photon
emission can be exclusively simulated. 

The generation of the photon's momenta has to be carefully
considered. As suggested by eq. \ref{APitera}, when the PS generates photons,
their energy is extracted from  the Altarelli-Parisi vertex $P(x)$, where $x$
is the energy fraction remaining to the fermion after the photon emission.

In order to generate also the angular variables, different solutions can be
exploited \cite{myplb}, being those degrees of freedom  integrated out in eqs.
\ref{qeddglap} and \ref{APitera}. The first possible choice is to weight
the photon angles according to the leading poles which are present in the
radiation matrix element. Namely, the photons' angles can be distributed 
according to the following formula
\begin{equation}
\cos\vartheta_\gamma\propto\sum_{i=1,4}\frac{1}{p_i\cdot k}
\label{leadingpole}
\end{equation}
where $p_i$ are the momenta of the charged particles and $k$ is the momentum
of the generic photon. If eq. \ref{leadingpole} can be a first
approximation, it is not so accurate when photonic variables are looked
at closer. In order to improve the exclusive photon generation, we can
observe that in eq. \ref{leadingpole} the interference effects between ISR
and FSR are completely neglected, thus the fermions radiate independently.

The radiation coherence effects can be included if we consider the cross
section for the production of $n$ photons from a generic process involving
fermions as external particles. Taking 
the soft limit and keeping all the interference terms, the resulting formula is
\begin{equation}
d\sigma_n\approx d\sigma_0\frac{1}{n!}\prod_{l=1}^n\frac{e^2d^{\it 3}
\mathbf{k}_l}{(2\pi)^32k^0_l}\sum_{i,j}\frac{\eta_i\eta_j p_i\cdot p_j}
{(p_i\cdot k_l)(p_j\cdot k_l)}
\label{coherence}
\end{equation}
where $k_l$ and $p_j$ are photons and fermions momenta and 
$\eta_i=\pm 1$ is a charge factor. The angular spectrum of the photons can be
directly extracted from the eq.
\ref{coherence}. This highly enhances the reliability of the PS also in the
simulation of radiative events, as will be shown in section \ref{pheno}.

An important point is the estimate of the physical precision of the PS
approach. In order to give a consistent estimate, an $\cal{O}(\alpha)$ PS has
been developed as well, which calculates the corrected cross section of eq.
\ref{sezfs} at first order in $\alpha$ in the PS approximation. By a systematic
comparison of the
the $\cal{O}(\alpha)$ PS results to the exact $\cal{O}(\alpha)$ matrix element
ones, the PS precision can be estimated and, in particular, the size of the
$\cal{O}(\alpha)$ non-leading-log contributions missing in a pure PS approach
can be quantified.

\subsection{The hard scattering cross sections}
\label{HSxsect}
The other ingredient entering eq. \ref{sezfs} is the ``kernel'' cross section,
which is the Born-like cross section for the process evaluated at the reduced
center of mass energy $x_1x_2s$. In $\tt BABAYAGA$, the standard formulae for
Bhabha, $\gamma\gamma$ and $\mu^+\mu^-$ processes are implemented. For the 
$\pi^+\pi^-$ channel, the pion Form Factor as parameterized in
\cite{khuemsantamaria} is present. The hard scattering cross section includes
also vacumm polarization contributions. It is worth discussing in some 
details how they are
treated in the last official release of $\tt BABAYAGA$ ($\tt 3.5$).

The running of $\alpha(q^2)$ is given by
\begin{eqnarray}
&&\alpha(q^2)=\frac{\alpha (0)}{1-\Delta\alpha (q^2)}\nonumber\\
&&\Delta\alpha (q^2)=\Delta\alpha_{had}^{(5)}(q^2)+\Delta\alpha_{t}(q^2)
+\nonumber\\
&&+\sum_l\Delta\alpha_{l}(q^2)\label{alphaq2}\end{eqnarray}
The lepton and top quark contributions ($\Delta\alpha_{l}(q^2)$ and 
$\Delta\alpha_{t}(q^2)$) do not show any problem, being analytically calculable
for any $q^2$ value.

The hadronic contribution $\Delta\alpha_{had}^{(5)}(q^2)$ is calculated by
means of the subroutine $\tt HADR5$ \cite{hadr5} (which is now superseded by
a new version).
This subroutine is not suited to work in the $q^2$ time-like region below 
($40$ GeV)$^2$. As suggested by one of the authors \cite{priv}, the following
solution was therefore adopted: in this $q^2$ region, the $q^2$ is flipped
to $-q^2$, in order to include at least the leading terms of the
hadronic contribution to $\Delta\alpha(q^2)$.

Recently, a new version of the routine has been written \cite{hadr5new} with
an improved treatment of the resonances region (including new experimental
data) and which is suited for
any $q^2$ value. It has been preliminarly included in the program, showing 
a difference on the integrated Bhabha cross section at DA$\Phi$NE energies 
at a few per mille level. However, a deeper look is needed and will be carried
out in the next future.

\section{The benchmark calculation $\tt LABSPV$}
\label{LABSPV}
One of the crucial points to be considered is the estimate of the theoretical
accuracy of the PS approach. In order to quantify the theoretical error, 
the independent program $\tt LABSPV$ has been derived from the code
$\tt SABSPV$ \cite{sabspv}, which was extensively used for LEP luminometry. It
is not a true event generator, but it is a cross section integrator.

In $\tt LABSPV$, the exact order $\alpha$ matrix element is used, and
the $\cal{O}(\alpha)$-corrected cross section is calculated according to
the standard formula
\begin{equation}
\sigma_{exact}^\alpha=\sigma_{S+V}^\alpha(E_\gamma<k_0)+
\sigma_{H}^{\alpha}(E_\gamma > k_0)\label{sezex}
\end{equation}
where $\sigma^\alpha_{S+V}$ is the soft-photon plus virtual cross section,
$\sigma^\alpha_H$ is the hard-photon emission cross section and $k_0$ is a
soft-hard radiation separator, which the total cross section does not depend
on. The code $\tt LABSPV$ calculates the cross sections for the Bhabha, 
$\gamma\gamma$ and $\mu^+\mu^-$ (only for high energies, $E_{c.m.}\gg m_\mu$)
processes, using matrix elements and formulae which can be found
in \cite{mesv,gg,mehard}. In the Bhabha and $\mu^+\mu^-$ case, vacuum 
polarization is treated as done in $\tt BABAYAGA$.

On top of the exact $\cal{O}(\alpha)$ cross section of eq. \ref{sezex}, in $\tt
LABSPV$, higher order leading-log QED corrections in a SF
approach can be also included, according to the relation
\begin{equation}
\sigma^{h.o.} = \sigma_{exact}^\alpha - \sigma_{SF}^\alpha + \sigma_{SF}^\infty
\label{sigmaadd}
\end{equation}
where $\sigma_{SF}^\alpha$ is the order $\alpha$ expansion of eq. \ref{sezfs}
and $\sigma_{SF}^\infty$ is eq. \ref{sezfs}. In $\tt LABSPV$, the SF's
are an analytical solution of the QED DGLAP equation \cite{labspvsf}
and are strictly collinear.
Equation \ref{sigmaadd} can be further improved if cast in the factorized form
\begin{equation}
\sigma^{h.o.} =\sigma_{SF}^\infty (1+\frac{\sigma_{exact}^\alpha -
 \sigma_{SF}^\alpha}{\sigma_0})
\label{sigmafact}
\end{equation}
It can be shown \cite{a2L} that in expression \ref{sigmafact} the bulk of the
$\alpha^2L$ corrections ($L$ is the collinear logarithm) are 
also taken into account, in addition to the exact order $\alpha$ and
leading-log higher order contributions. As a consequence, the expected
theoretical accuracy of this approach for cross section calculation can be
estimated at the level of 0.1\%.
\begin{table}[tb]
\caption{Comparison at different center of mass energies of
the total $\gamma\gamma$ cross section between $\tt LABSPV$ and the results
of ref. \cite{gg} (B. \& K.)}
\newcommand{\m}{\hphantom{$-$}}
\newcommand{\cc}[1]{\multicolumn{1}{c}{#1}}
\renewcommand{\tabcolsep}{2pc} 
\renewcommand{\arraystretch}{1.2} 
$$\begin{array}{l|l|l}\hline
\multicolumn{3}{c}{\mbox{$e^+e^-\to\gamma\gamma(\gamma)$ total cross
sections}}\\
\hline
\mbox{energy (GeV)}&\mbox{B. \& K.}&\mbox{{$\tt LABSPV$}}\\[1mm]
\hline
 1 & 2013.7 & 2013.6(3)\\
 3 & 264.90 & 264.92(5)\\
 5 & 102.58 & 102.59(2)\\
 7 & 54.82  &  54.82(1)\\
10 & 28.184 &  28.182(6)\\
\hline\end{array}$$\label{table1}\end{table}

Being used as a ``benchmark'', $\tt LABSPV$ has been compared to different
independent calculations finding very good agreement. For example,
in table \ref{table1}, the total cross section for $\gamma\gamma$ is compared
to the analytical results of \cite{gg}.
\section{Phenomenological results}
\label{pheno}
\subsection{Coherent Parton Shower}
As discussed in paragraph \ref{PSalgo}, the photon angular generation can
be performed in order to simulate also the ISR-FSR interference, which is
neglected in a pure ``leading-pole'' generation. The coherence effect can be
important when exclusive photon distributions are considered.
\begin{figure}[tb]
\includegraphics[height=8cm]{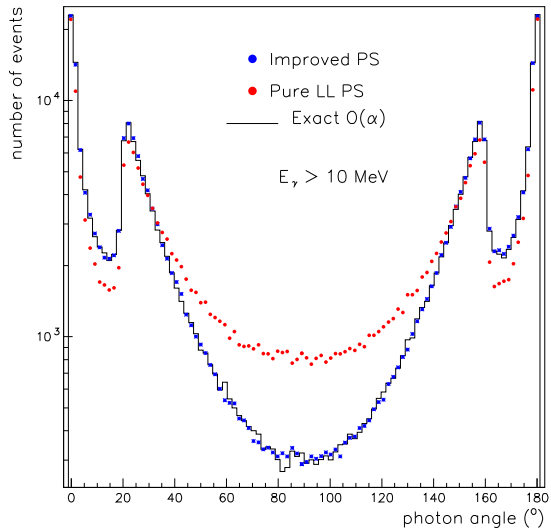}
\caption{The photon angular distribution as obtained by the leading-pole
$\cal{O}(\alpha)$ PS
(red dots), the coherent $\cal{O}(\alpha)$ PS (blue dots) and the exact 
$\cal{O}(\alpha)$ matrix element (histogram). Cuts are given in the text.}
\label{cohe1}\end{figure}

\begin{figure}[tb]
\includegraphics[height=8cm]{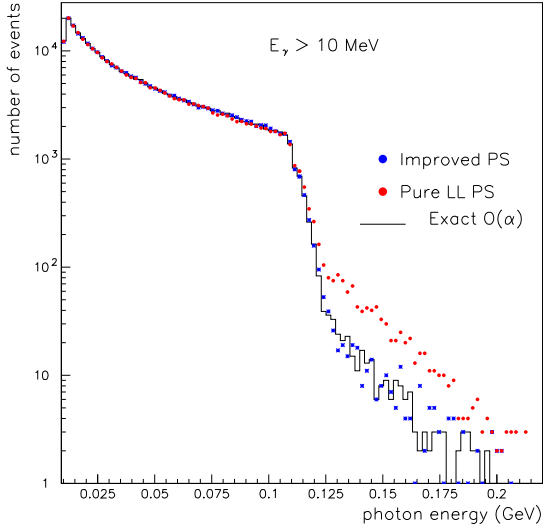}
\caption{The photon energy distribution as obtained by the leading-pole 
$\cal{O}(\alpha)$ PS
(red dots), the coherent $\cal{O}(\alpha)$ PS (blue dots) and the exact
$\cal{O}(\alpha)$ matrix element (histogram). Cuts are given in the text.}
\label{cohe2}\end{figure}

In fig. \ref{cohe1}, the photon angular distribution is shown for
$e^+e^-\to e^+e^-\gamma$ events at $\sqrt{s}=1.02$ GeV, with
$20^\circ<\vartheta_\pm<160^\circ$, $E_\pm>0.4$ GeV and $E_\gamma>10$ MeV.
The leading-pole and coherent $\cal{O}(\alpha)$ PS results obtained with
$\tt BABAYAGA$ are compared to the exact $\cal{O}(\alpha)$ distribution of
$\tt LABSPV$. The improvement of the coherence PS is evident.
In fig. \ref{cohe2}, the distribution of the photon energy is plotted for
the same sample of events. Even if the leading-pole PS already produces a good
distribution, being the photon energy driven by the Altarelli-Parisi vertex,
the coherent PS is better also in the tail of the energy spectrum.

\begin{figure}[tb]
\includegraphics[height=8cm]{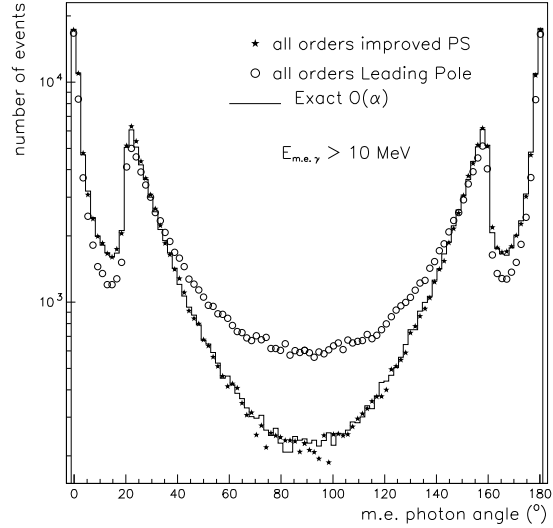}
\caption{The most energetic photon  distribution as obtained by the
leading-pole PS (open circles) and the coherent PS at all orders (stars)
compared exact to the $\cal{O}(\alpha)$  matrix element 
photon distribution (histogram).}\label{cohe_exp}\end{figure}

A similar behaviour is found for the PS at all orders: in fig. \ref{cohe_exp},
the angular distribution of the most energetic photon as obtained by the
coherent and the leading-pole PS at all orders are compared to the exact
$\cal{O}(\alpha)$ distribution (even if this is not fully consistent).
\subsection{Theoretical error}
The theoretical error of $\tt BABAYAGA$ for cross section calculation can be
evaluated by comparing the $\cal{O}(\alpha)$ PS to the exact matrix element of
eq. \ref{sezex}. Actually, the main contribution to the error comes from the
non-leading-log $\cal{O}(\alpha)$ terms, missing in the PS approach.

\begin{figure}[tb]
\includegraphics[height=4cm,width=8cm]{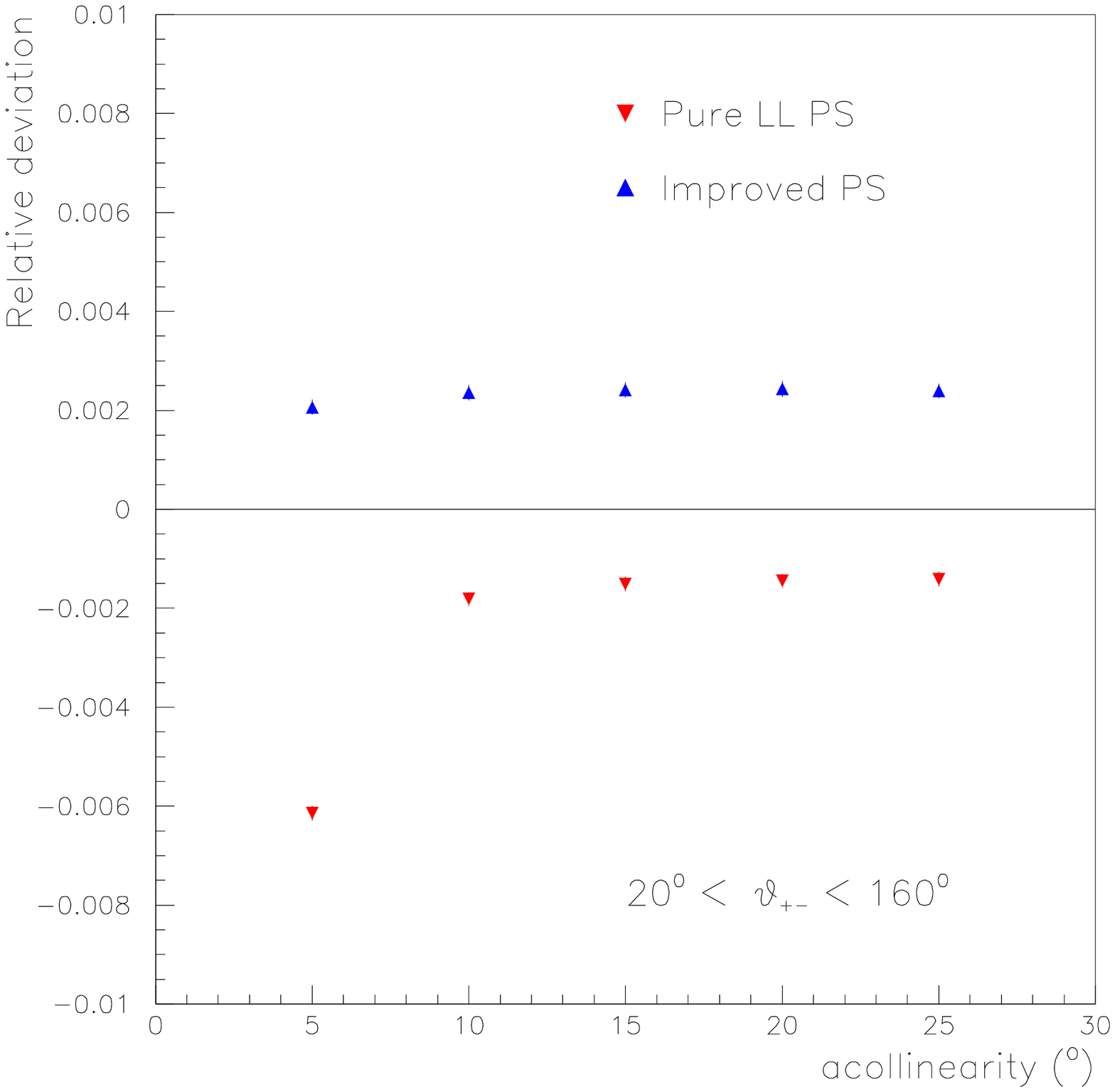}
\includegraphics[height=4cm,width=8cm]{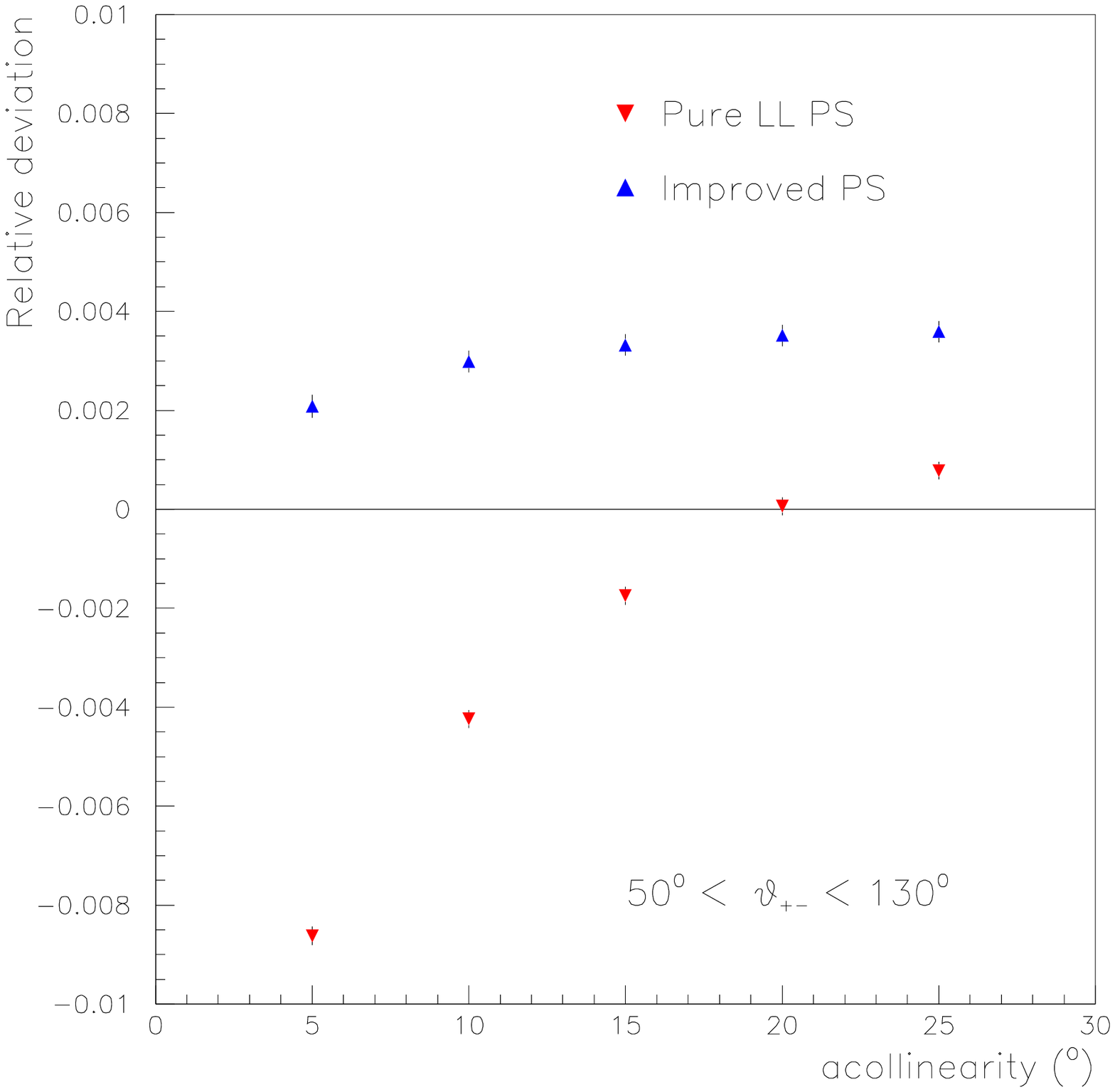}
\caption{Relative effects of missing $\cal{O}(\alpha)$ contributions in
$\tt BABAYAGA$. Blue triangles represent the coherent PS and the red ones
the leading-pole PS. See text for details.}\label{missoa}\end{figure}

In fig. \ref{missoa}, the relative deviation between the inclusive 
Bhabha cross section obtained with $\tt BABAYAGA$ at $\cal{O}(\alpha)$ and
the exact matrix element is plotted as a function of the acollinearity cut,
both for the leading-pole PS and the coherent PS. The imposed experimental
cuts are $\sqrt{s}=1.019$ GeV, $20^\circ<\vartheta_\pm<160^\circ$ (top in the
figure) or $50^\circ<\vartheta_\pm<130^\circ$ (bottom in the figure),
$E^{min}_\pm=0.4$ GeV.

From fig. \ref{missoa}, it emerges that the error in $\tt BABAYAGA$ is at
level of $0.5\%$ or lower. It is worth noticing that for the coherent PS the
size of the missing $\cal{O}(\alpha)$ is almost constant as the cuts vary,
while for the leading-pole PS it is not constant and it can become larger than
$0.5\%$.

For the Bhabha case, we can fix the theoretical error of $\tt BABAYAGA$ in the 
calculation of the QED corrected cross section at the level of $0.5\%$.

\subsection{Higher order corrections}
Aiming at a high precision calculation of the QED processes cross section,
higher order contributions can not be neglected, expecially if a ``bare'' event
selection criterium is adopted and almost-elastic events are selected.

\begin{figure}[tb]
\includegraphics[height=4cm,width=8cm]{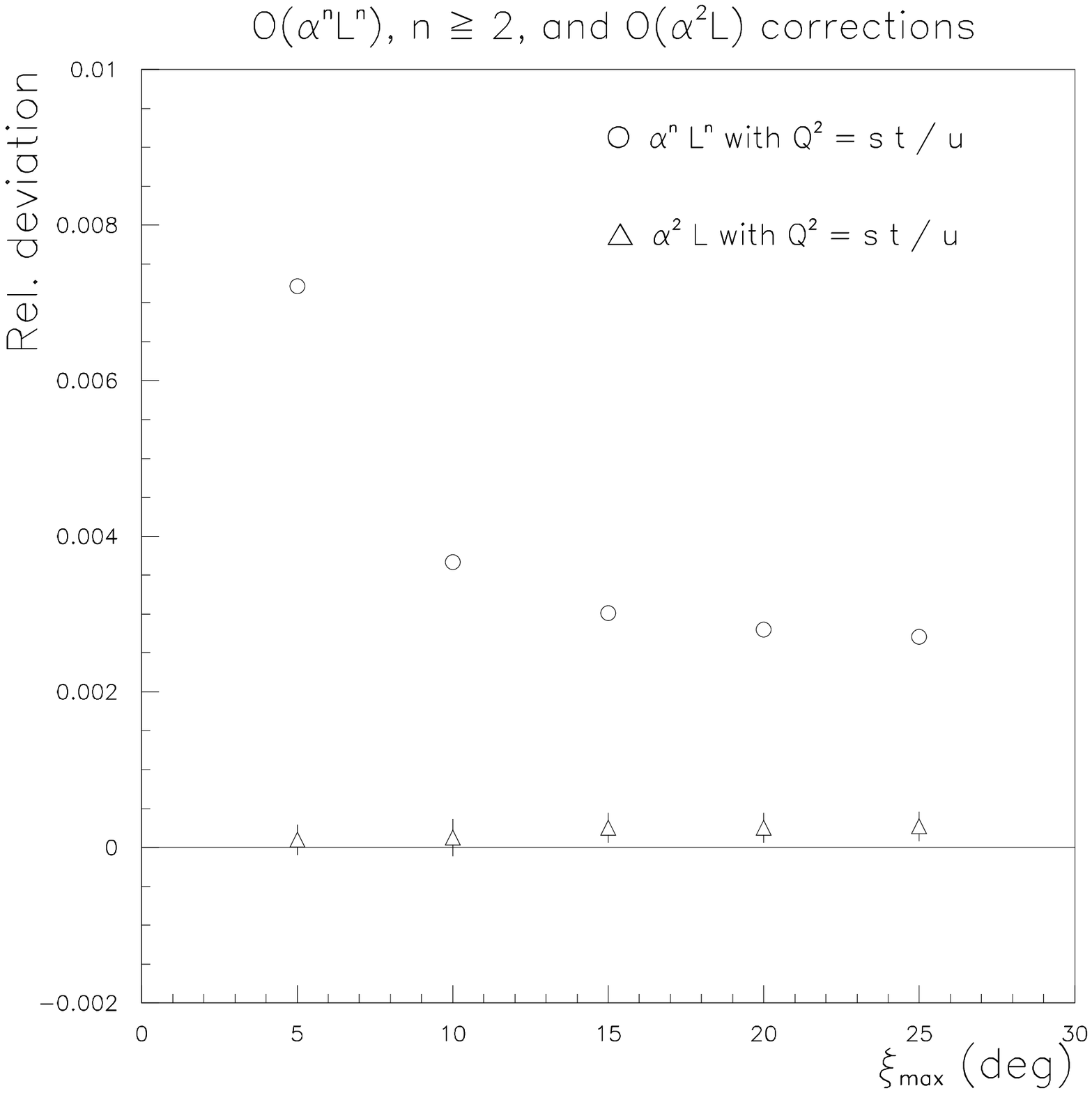}
\includegraphics[height=4cm,width=8cm]{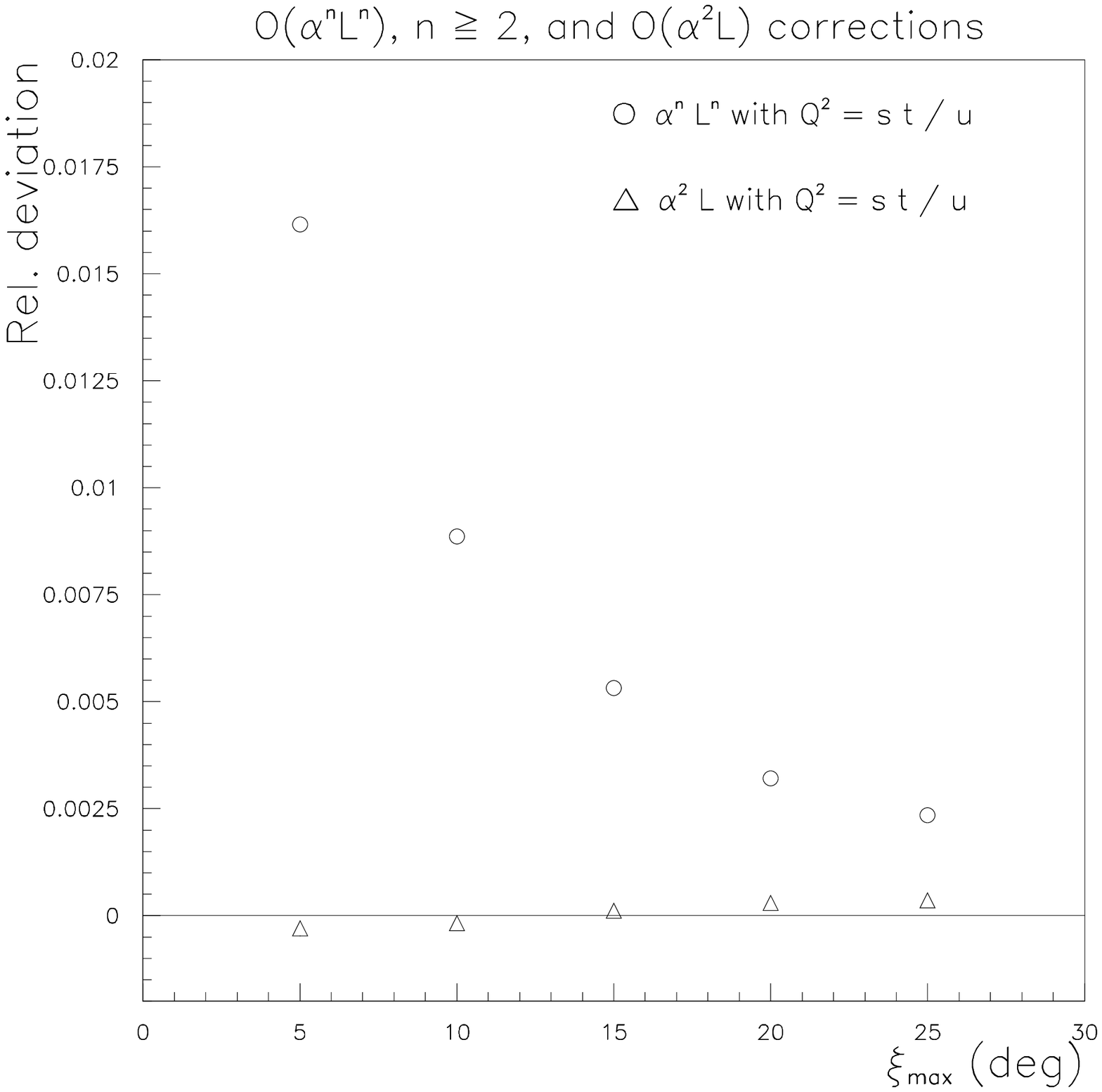}
\caption{Relative effects of higher-order contributions obtained with
$\tt BABAYAGA$ and $\alpha^2L$ contribution obtained with $\tt LABSPV$.
See text for details.}\label{hocontrib}\end{figure}

The PS allows to include higher-order corrections in a natural way. In
fig. \ref{hocontrib}, the relative difference between the cross sections
obtained with $\tt BABAYAGA$ at $\cal{O}(\alpha)$ and at all orders is shown.
The simulation has been performed applying the same cuts as in fig.
\ref{missoa}. As it can be seen, the higher-order contributions are quite
large: they can change the cross section for an amount of $1-2\%$, depending
on the cuts, and are therefore important to achieve the desired theoretical
accuracy. The higher-order impact can be also estimated with $\tt LABSPV$,
finding a good agreement. In the plots of figure \ref{hocontrib},
the effect of the $\alpha^2L$ contributions is also shown,
estimated with $\tt LABSPV$. They are negligible, being below the $5\cdot
10^{-4}$.

\begin{figure}[tb]
\includegraphics[height=8cm,width=8cm]{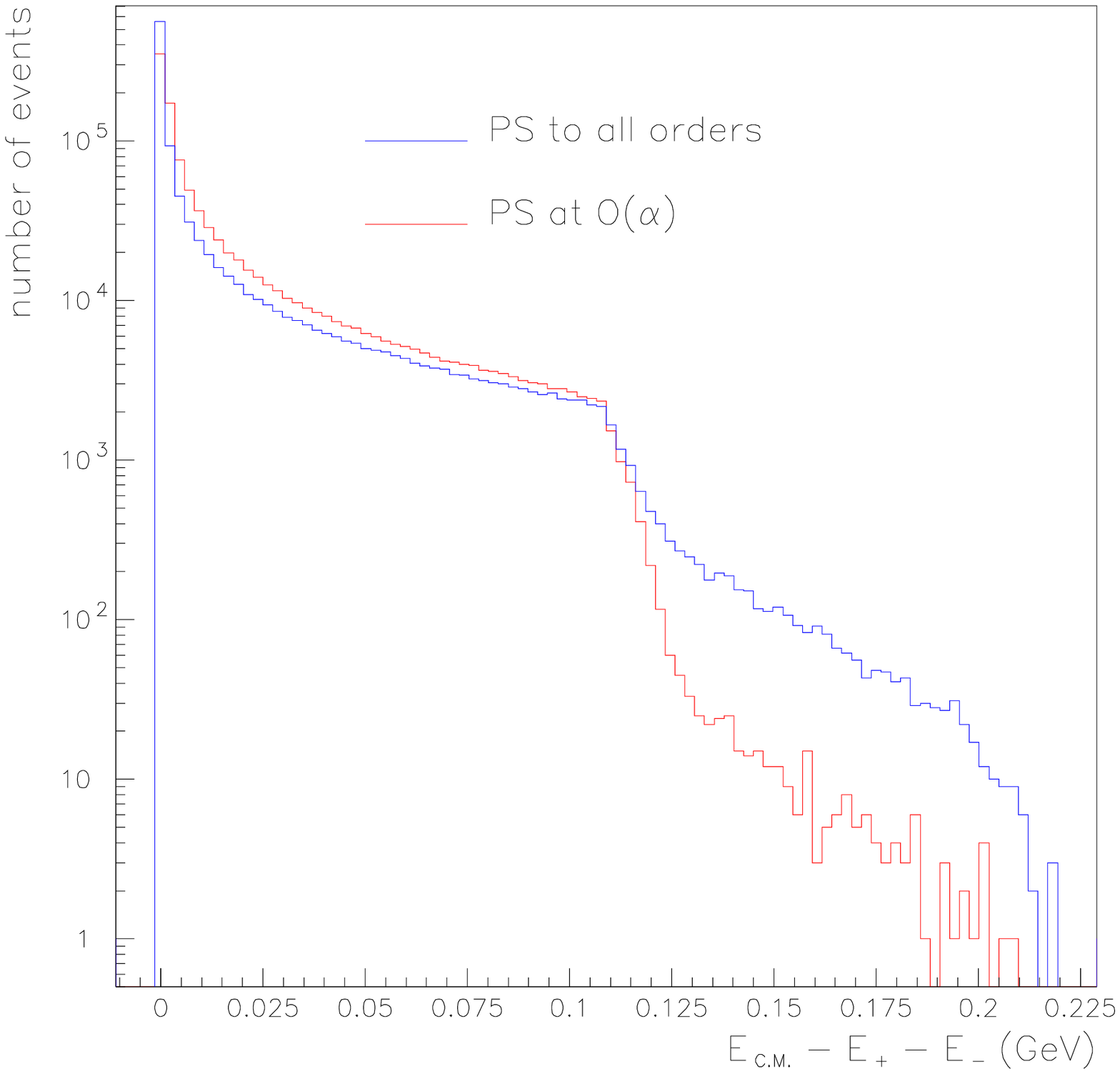}
\caption{Effect of multi-photon emission on the radiation energy
distribution}\label{missenergy}\end{figure}

It interesting to notice that multi-photon emission has a sizeable effect
also on distributions. For example, in figure \ref{missenergy} the
distribution of radiation energy, defined as $E_{c.m.}-E_+-E_-$, is plotted
and the $\cal{O}(\alpha)$ PS is compared to the all order PS.


\section{Other channels}
$\tt BABAYAGA$ can generate also $\mu^+\mu^-(n\gamma)$, $\gamma\gamma(n\gamma)$
and $\pi^+\pi^-(n\gamma)$ final states, generating events according
to eq. \ref{sezfs} and the PS algorithm.

For $\mu^+\mu^-$ and $\gamma\gamma$ processes, a study as detailed as for
Bhabha process to establish the theoretical accuracy has not been performed
yet, but it is planned. 

As a general consideration, it should be noticed that in
the benchmark program $\tt LABSPV$ the fully massive $\mu^+\mu^-(\gamma)$ exact
$\cal{O}(\alpha)$ matrix element is not present and therefore it is not
reliable at low energies, where the muon mass effects become important.

Concerning the $\gamma\gamma$ final state, it has to be pointed out that
the PS approach suffers for double counting. In fact, the photons generated
by the PS are independent from the photons of the hard scattering process:
if the PS photons are allowed to go inside the phase space region of the hard
scattering ones, this leads to a double counting error. However, the error
can be kept small if large angle photons are selected, because PS emits photons
preferably collinear to initial-state $e^+$ and $e^-$.

In the version $\tt 3.5$ of the program, also the $\pi^+\pi^-(n\gamma)$
channel has been added. In this case, only ISR is simulated with the PS. With
the PS approach, also the radiative $\pi^+\pi^-\gamma$ event can be studied, 
which is extremely important for the $R$ measurement via the radiative return
method. The process was added only for cross-checking purposes, being the $\tt
PHOKHARA$ MC event generator \cite{phokhara} much more accurate for
$\pi^+\pi^-\gamma$.

\begin{figure}[tb]
\includegraphics[height=8cm,width=8cm]{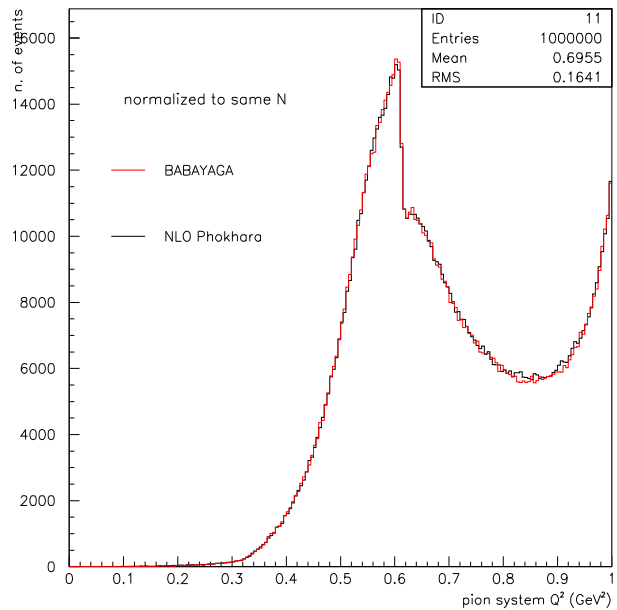}
\caption{$\pi^+\pi^-$ invariant mass distribution by $\tt BABAYAGA$ and $\tt
PHOKHARA \ 1.0$. Cuts are given in the text.}\label{pi+pi-}\end{figure}

Nevertheless, the results
of $\tt BABAYAGA$ are quite good. In figure \ref{pi+pi-}, the $\pi^+\pi^-$
invariant mass distribution obtained by $\tt BABAYAGA$ is compared to
the $\tt PHOKARA \ 1.0$ distribution. The imposed cuts are $\sqrt{s}=1.02$ GeV,
$55^\circ<\vartheta_\pm<125^\circ$, $5^\circ<\vartheta_\gamma<21^\circ$ and 
$E_\gamma>10$ MeV.

\section{Towards a new release of $\tt BABAYAGA$}
\begin{table*}[htb]
\caption{Preliminary comparisons for Bhabha cross sections (in $nb$) obtained
with $\tt LABSPV$, $\tt BABAYAGA \ 3.5$ and an improved version of $\tt
BABAYAGA$. Cuts are given in the text.}
\renewcommand{\tabcolsep}{2pc}
\renewcommand{\arraystretch}{1.2}
$$\begin{array}{l|l|c|c|c|c}
\mbox{acceptance} & \mbox{RC} &\mbox{$\tt LABSPV$ (eq. \ref{sigmaadd})}&\mbox{$\tt LABSPV$ (eq. \ref{sigmafact})}&\mbox{{$\tt BABAYAGA \ 3.5$}}&
\mbox{improved $\tt BABAYAGA$}\\
\hline
55^\circ - 125^\circ&\mbox{$\cal{O}(\alpha)$}& 451.70(1) & & 454.76(1) & 451.75(1)\\
&\mbox{$\cal{O}(\alpha)$+h.o.}  & 456.20(2) & 456.36(4)& 458.86(1) & 455.96(2)\\\hline
20^\circ - 160 ^\circ&\mbox{$\cal{O}(\alpha)$} & 6061.9(2) & & 6088.8(1) & 6062.5(2)\\
&\mbox{$\cal{O}(\alpha)$+h.o.}  & 6087.1(3) & 6090.0(7)&6114.4(1) & 6088.2(3)\\
\hline\end{array}$$\label{table2}\end{table*}

As discussed in the section \ref{pheno}, the main source of error in a PS
approach comes from the missing $\cal{O}(\alpha)$ non-log terms. It would
be extremely useful to include them in the generator, in order to further
reduce its theoretical error. In the measurements of the KLOE collaboration
for the $R$ ratio, the theoretical error in the luminosity is going to be the
dominant one \cite{incagli}.

When trying to include the exact $\cal{O}(\alpha)$ matrix element in the PS for
exclusive event generation~\footnote{This is an extensively studied
topic also in QCD.
See for instance \cite{frixweb}.}, different merely technical problems arise
(negative weigths, higher-order not well under control, etc.). Work in this
direction is in progress. 

Here, we would like to show some preliminary results obtained with a new
version of $\tt BABAYAGA$, where part of the exact $\cal{O}(\alpha)$ matrix
element for Bhabha scattering is used.

Roughly, the idea is the following: the events are reweigthed for a
(infrared-independent) factor which accounts for the difference between the
exact and the PS soft-plus-virtual cross section. Radiative events should then
be reweigthed with the exact $\cal{O}(\alpha)$ radiative matrix element. At 
present, we did not succeed to perform exactly the latter step. Nevertheless,
it is possible to find out from eq. \ref{coherence} a weight factor which is 
much closer to the exact matrix element than what is implemented in
$\tt BABAYAGA \ 3.5$. The technical details will be discussed elsewhere.

It is worth stressing that this procedure is done at differential level and
the order $\alpha$ expansion of the resulting cross section reproduces
exactly the soft-plus-virtual cross section ($\sigma^\alpha_{S+V}$) of
eq. \ref{sezex} and ``almost'' exactly the hard photon term 
($\sigma^\alpha_H$) of eq. \ref{sezex}.

In table \ref{table2}, the cross section of $\tt LABSPV$, $\tt BABAYAGA \ 3.5$ 
and new version are compared. The applied cuts 
for the cross sections shown in the table
are $\sqrt{s}=1.02$ GeV, $20^\circ<\vartheta_\pm<160^\circ$ 
or $55^\circ<\vartheta_\pm<125^\circ$, $E_\pm^{min}>0.408$ GeV, maximum
acollinearity $\zeta_{max}=10^\circ$. In table \ref{table2}, the vacuum
polarization contributions have been switched off. The differences between the
improved version of $\tt BABAYAGA$ and $\tt LABSPV$ are below the 0.1\%.

\section{Conclusions}
The program $\tt BABAYAGA$ is a MC event generator for QED processes and
$\pi^+\pi^-$ final state at flavour factories. The QED radiative corrections
are included by means of a Parton Shower. The theoretical accuracy of the
approach is estimated to be at $0.5\%$ for cross section calculation.

Some preliminary results have been presented of a new
version of the generator, which is under development. In the next future, it
will include also
the exact $\cal{O}(\alpha)$ matrix element and will improve the theoretical 
accuracy of the approach.

\section{Acknowledgments}
The authors would like to thank the organizers of the SIGHAD03 Workshop for
the pleasant athmosphere and the useful discussions during the workshop.

\end{document}